\documentclass[preprint,5p,twocolumn]{elsarticle}

\usepackage[colorlinks = true,linkcolor = blue,urlcolor  = blue, citecolor = blue]{hyperref}
\usepackage{graphicx}
\usepackage{siunitx}
\usepackage{subcaption}
\usepackage{booktabs}
\usepackage{amsmath}
\usepackage{textcomp}
\usepackage{footnote}

\usepackage{amssymb}
\usepackage[version=4]{mhchem}

\usepackage[capitalise]{cleveref}

\journal{Nuclear Fusion}

\begin{document}

\begin{frontmatter}

\title{BABY 1L: First Tritium Breeding Campaign Results}

\cortext[cor1]{Corresponding author}

\author[MIT]{Rémi Delaporte-Mathurin\corref{cor1}}%
\ead{remidm@mit.edu}

\author[MIT]{Nikola Goles}
\author[MIT]{Collin Dunn}
\author[MIT]{Emily Edwards}
\author[MIT]{Sara Ferry}
\author[UKAEA]{Ross MacDonald}
\author[MIT]{Ethan Peterson}
\author[MIT,PoliTO]{Davide Pettinari}
\author[MIT]{Stefano Segantin}
\author[MIT]{Weiyue Zhou}
\author[MIT]{Kevin B. Woller}

\address[MIT]{Plasma Science and Fusion Center, Massachusetts Institute of Technology, Cambridge, MA 02139, USA}
\address[UKAEA]{United Kingdom Atomic Energy Authority, Culham Campus, Abingdon, Oxfordshire, OX14 3DB, UK}
\address[PoliTO]{Department of Energy “Galileo Ferraris”, Politecnico di Torino, Corso Duca degli Abruzzi 24, Torino, 10129, Italy}

\begin{abstract}
Achieving tritium self-sufficiency is a critical challenge for future fusion power plants.
The BABY 1L experiment, part of the LIBRA project at MIT, aims to benchmark tritium breeding and release in molten salt breeder systems under deuterium-tritium (DT) neutron irradiation.
Building on the initial \SI{100}{mL} campaign, BABY 1L introduces a tenfold increase in breeder volume, improved thermal and gas handling systems, and enhanced neutron diagnostics, including a proton recoil telescope.
We report on results from four irradiation experiments using sealed-tube DT neutron generators, with tritium collected by water bubblers measured via liquid scintillation counting.
Experimentally determined Tritium Breeding Ratios (TBRs) were compared to OpenMC neutronics simulations, showing very good agreement.
The measured TBR values demonstrate a six-fold improvement over the \SI{100}{mL} experiments, largely attributed to the increased solid angle and improved measurement fidelity.
We also investigate tritium release dynamics and identify diffusion-limited transport as the dominant regime in the salt volume in the temperature range 630-750 \si{\celsius}.
Additionally, we observe that the introduction of hydrogen in the helium carrier gas significantly accelerates tritium release, consistent with an isotopic exchange mechanism.
All analysis is conducted through the open-source \texttt{libra-toolbox} \cite{libra-toolbox}, which streamlines simulation, data processing, and validation across experimental campaigns.
These results provide critical insights into the design and operation of future liquid breeder systems and demonstrate the maturity of the BABY platform as a testbed for tritium breeding studies.

\end{abstract}



\end{frontmatter}


\section{Introduction}

Tritium self-sufficiency is a fundamental requirement for the viability of future fusion power plants, as tritium is radioactive, rare in nature, and cannot be generated at the quantities necessary for GW-scale fusion power use.
Liquid breeder systems, such as those based on molten salts, offer a promising pathway for achieving this goal due to their high lithium densities and good heat transfer properties \cite{sorbom_arc_2015}.
However, the understanding of tritium generation and transport in such systems remains limited, particularly under relevant DT neutron conditions. 

The BABY experiment was designed to fill this knowledge gap by enabling well-controlled measurements of tritium breeding and release in molten salt breeder concepts under \SI{14}{MeV} neutron irradiation.
Building upon the initial \SI{100}{mL} campaigns \cite{delaporte-mathurin_advancing_2025}, the BABY 1L upgrade introduced a tenfold increase in salt volume and a suite of experimental improvements.
These include enhanced temperature control, refined gas handling, added sparging capabilities, and real-time neutron diagnostics.

The primary goal of the BABY 1L campaign is to quantify the Tritium Breeding Ratio (TBR) and tritium release dynamics under different operational conditions, and to compare these measurements to model predictions.
The campaign combines neutron activation diagnostics, real-time neutron flux monitoring, and liquid scintillation counting with detailed neutronics and transport simulations using tools such as OpenMC.

In this paper, we present results from the first four BABY 1L experimental runs. \Cref{methodology} describes the experimental setup, instrumentation, and modelling tools.
\Cref{results} focuses on neutron flux measurements, TBR calculation, tritium release behaviour, and tritium speciation.
We place special emphasis on the effect of hydrogen addition to the sweep/sparge gas and its impact on release dynamics.
\Cref{discussion} provides a critical assessment of uncertainties, modelling limitations, and remarks on external cross-irradiation sources.

\section{Methodology}\label{methodology}
\subsection{Experiment}

\begin{figure*}[h]
    \centering
    \includegraphics[width=0.75\linewidth]{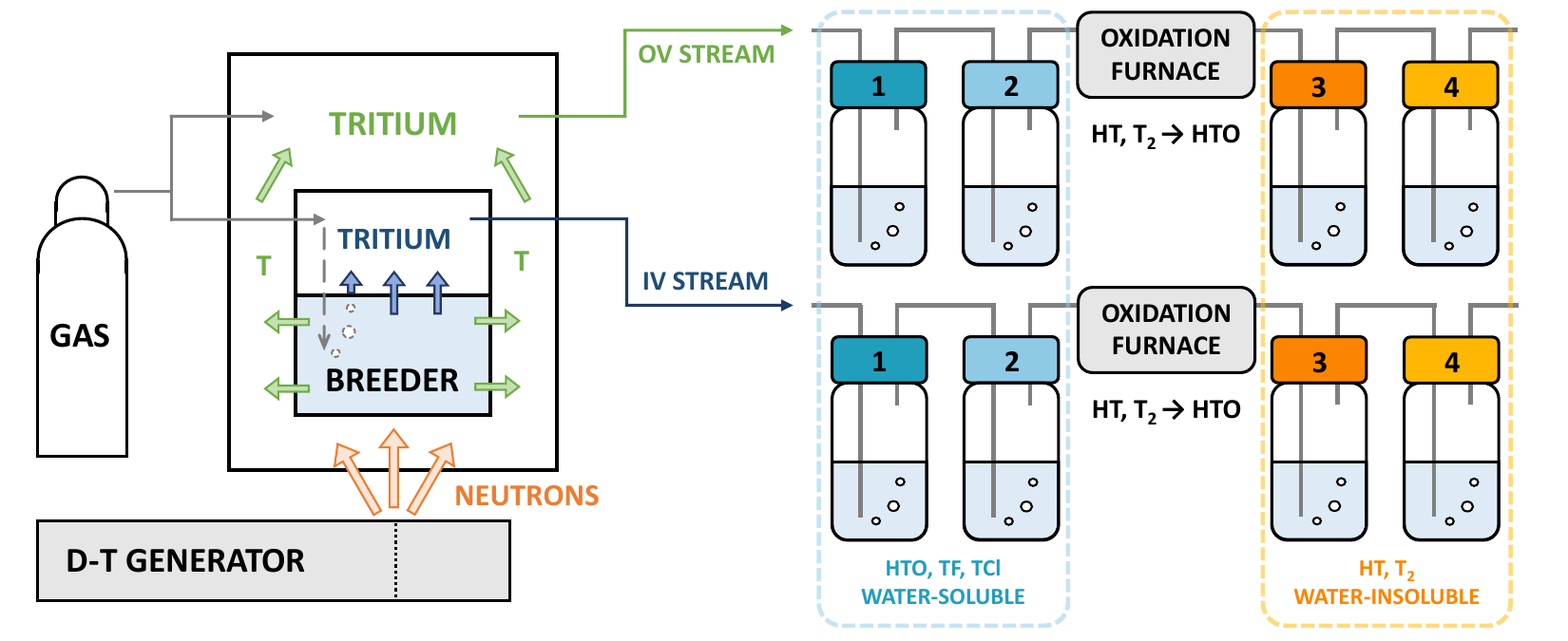}
    \caption{Simplified diagram of the experimental system, consisting of the vessel assembly, neutron generator, and gas handling system with bubblers for tritium collection.}
    \label{fig:simplified-diagram}
\end{figure*}

The experimental system  is centered around a crucible that contains 1L of the molten breeder salt (see Figure \ref{fig:simplified-diagram}).
The crucible is enclosed in an outer vessel which enables the collection of tritium that permeates through the walls of the crucible.
The salt is irradiated by a neutron source positioned outside of the vessel assembly, breeding tritium by the interaction between neutrons and the salt.
Once bred, the tritium transports through the salt before outgassing, either through the interface between the salt and the cover gas or by permeating through the walls of the crucible.
The gas, whose composition can be varied, is conditioned and flown at positive pressure and at a desired rate through two streams:
\begin{itemize}
    \item Inner Vessel (IV) stream, in which the gas passes through the crucible, either as cover gas over the salt or sparged through the melt, and collects the tritium released through the salt-gas interface;
    \item Outer Vessel (OV) stream, in which the gas sweeps through the outer vessel around the crucible and collects the permeated tritium.
\end{itemize}
A set of water bubblers is connected to the outlet of each of the streams to collect the tritium carried by the gas. The bubblers consist of four vials each, in two pairs separated by an oxidation furnace which enables the differentiation of water-soluble and water-insoluble molecules containing tritium. \par

\subsubsection{DT neutron generators}

The BABY experiment can utilise two sealed-tube neutron generators: a Thermo-Fisher™ A-325 and a Thermo-Fisher™ P-383.
These devices produce neutrons through deuterium-tritium (DT) fusion reactions.
Inside each generator, a target composed of metal loaded with deuterium and tritium is positioned near a high-voltage electrode.
When voltage is applied, it generates an electric field that accelerates deuterium ions towards the target.
As the ions penetrate the target material, they lose energy through Coulomb interactions, and a subset undergoes nuclear fusion with tritium atoms, yielding helium nuclei and neutrons with energies around \SI{14}{MeV}.

The sealed nature of the generators keeps the tritium securely enclosed, making the sources suitable for controlled neutron production in experimental settings. However, the neutron spectrum produced is not limited to a monoenergetic \SI{14}{MeV} output.
Some deuterium-deuterium (D-D) reactions also occur, generating neutrons around \SI{2}{MeV}.

BABY-1L runs 1, 3 and 4 used the A-325 generator and run 2 used the P-383 generator.

\subsubsection{Breeder material and vessel assembly}

1.88 kg of ClLiF salt, an eutectic mixture of 30.5 mol\% LiF - 69.5 mol\% LiCl identical to the breeder in the BABY \SI{100}{mL} experiment \cite{delaporte-mathurin_advancing_2025}, was placed in a cylindrical Inconel-625 crucible, 5.75" (\SI{14.605}{cm}) OD, 4.75" (\SI{12.065}{cm}) tall with walls \SI{3}{mm} thick.
99.9\% LiCl, anhydrous from Apollo Scientific Ltd. and 99.99\% (metals basis) LiF from Thermo Scientific were used for making the ClLiF salt. 
Two 1/2" (\SI{1.27}{cm}) OD IV stream gas tubes stemming from the top of the crucible were welded through a 1" (\SI{2.54}{cm}) thick 304L stainless steel 12" (\SI{30.48}{cm}) conflat blank serving as the lid of the 10.75" (\SI{27.305}{cm}) OD, 8.3" (\SI{21.082}{cm}) tall cylindrical outer vessel made of 4.19 mm thick 304L SS.
When assembled, the crucible was suspended inside of the outer vessel via the gas lines.
A 1/2" (\SI{1.27}{cm}) OD tube for OV stream gas inlet was welded through the lid, extending along the outside of the crucible wall and terminating near the bottom of the outer vessel.
The OV stream outlet tube did not extend below the lid. \par

Two ceramic insulation plates were placed inside the outer vessel, one 0.19" (\SI{0.4826}{cm}) thick lining the bottom, and the other 0.28" (\SI{0.7112}{cm}) thick covering the top of the crucible.
A gap separated the bottom of the crucible from the bottom alumina plate.
The side walls of the outer vessel were lined with with a layer of 6" wide, 1/8" (\SI{0.3175}{cm}) thick silica fabric.
The vessel assembly was placed on a 0.75" (\SI{1.905}{cm}) thick epoxy workbench of unknown composition, lined with two layers of kaowool insulation, each 1" thick before compression under the weight of the vessel. \par

An Inconel-625 re-entrant tube was welded to the top of the crucible and lined internally with a 1.42 mm thick copper sheath to assist with heat transfer.
A 5/8" (\SI{1.5875}{cm}) OD Incoloy-800 sheathed 1200 W Watlow FIREROD\textsuperscript{\textregistered{}} cartridge heater was inserted along the vessel axis, through the outer vessel lid and into the lined re-entrant tube, reaching a depth 0.35" (\SI{0.889}{cm}) above the bottom of the crucible.
Inside the outer vessel, the crucible was surrounded by a 7.5" (\SI{19.05}{cm}) ID, 6" (\SI{15.24}{cm}) tall, 1000 W radiative furnace custom-built by DS Fibertech Corporation, which included a 1.125" (\SI{2.8575}{cm}) thick ceramic insulation layer. \par 

The cylindrical neutron generator was positioned under the workbench with its axis parallel to the epoxy plate.
A 1/4" (\SI{0.635}{cm}) air gap separated the body of the generator from the plate. \par

An annotated diagram showing the layout of the system components is shown on Figure \ref{fig:vessel_components}.

\subsubsection{Heating system and temperature distribution}

The cartridge heater and furnace power supply was managed using separate PID controllers, based on the temperature measured by thermocouples embedded within each heating element.
Steady-state parameters of the heating system are detailed in Table \ref{tab:heaters}.\par

\begin{table*}[h!]
    \centering
    \caption{Operating parameters of the heating system in steady state.}
    \begin{tabular}{lrrrr} 
        \toprule
         & Setpoint temperature & Temperature reached & Rated power & PID max power output limit\\
        \midrule
        Heater & \SI{750}{\celsius} & \SI{750}{\celsius} & \SI{1200}{W} & 60\% \\
        Furnace & \SI{700}{\celsius} & \SI{610}{\celsius} & \SI{1000}{W} & 90\% \\
        \bottomrule
    \end{tabular}
    \label{tab:heaters}
\end{table*}

\begin{figure}[h!]
     \centering
     \begin{subfigure}[b]{\linewidth}
         \centering
         \includegraphics[width=0.85\linewidth]{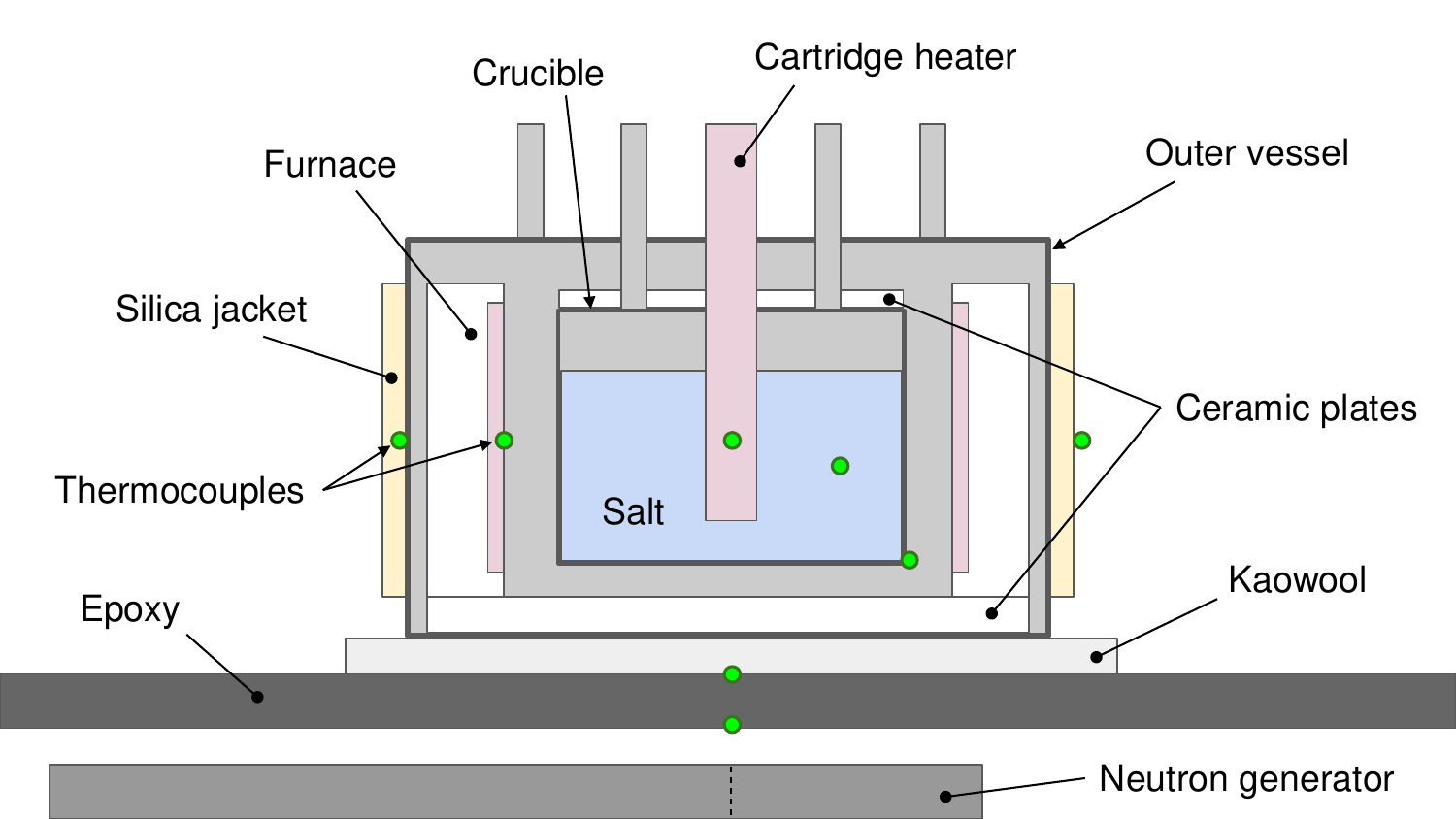}
         \caption{Major components of the vessel assembly. Blue area represents the salt, red - heating elements, white and light gray - solid and wool insulation, yellow - silica, gray - gas-filled volume, dark gray - epoxy, green dots - thermocouple sensor locations.}
         \label{fig:vessel_components}
     \end{subfigure}
     \begin{subfigure}[b]{\linewidth}
         \centering
         \includegraphics[width=0.85\linewidth]{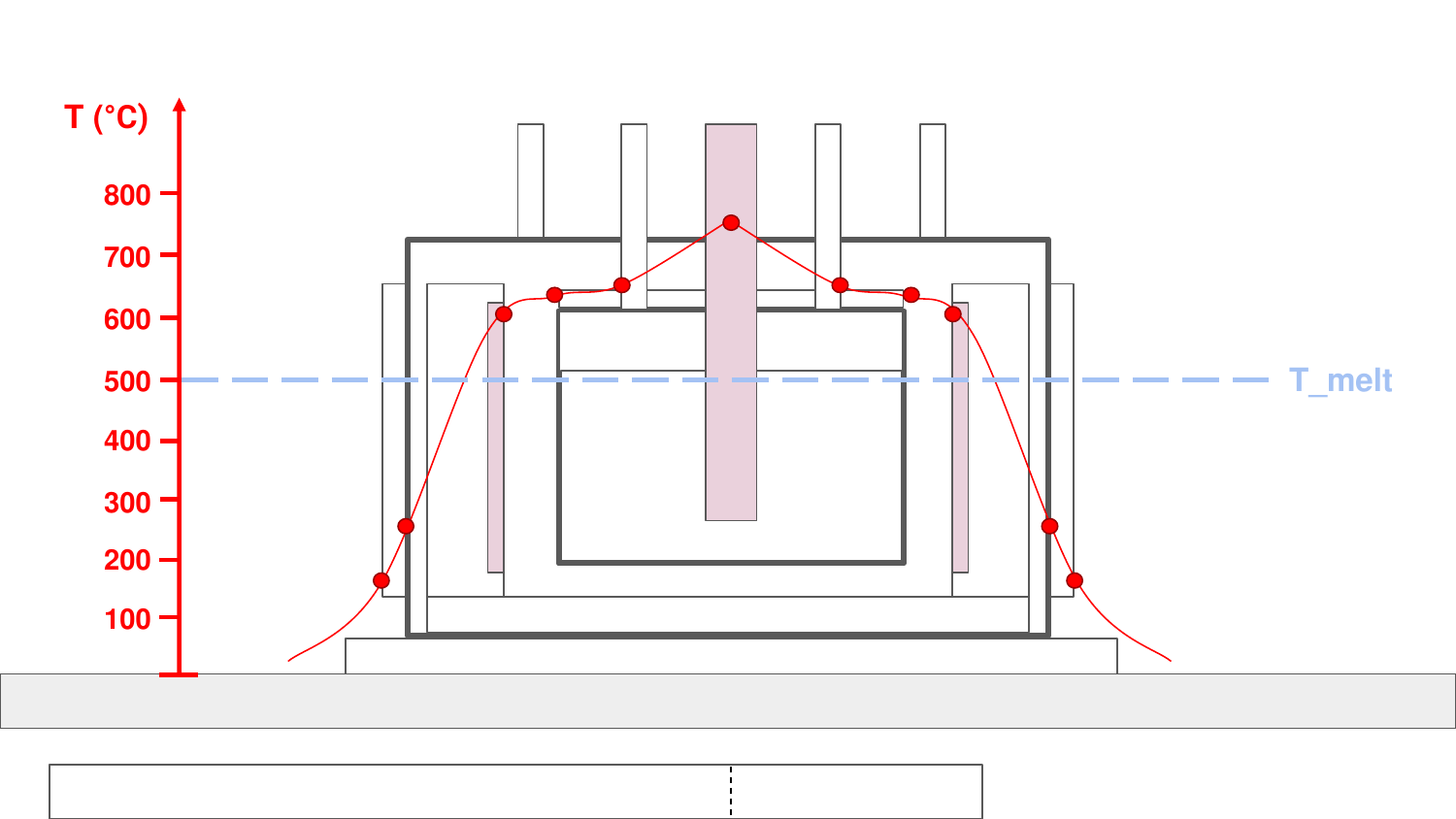}
         \caption{Approximate typical steady-state radial distribution of temperature, showing that the full salt volume is above the melting temperature of ClLiF.}
         \label{fig:radial_temperature_distribution}
     \end{subfigure}
     \begin{subfigure}[b]{\linewidth}
         \centering
         \includegraphics[width=0.85\linewidth]{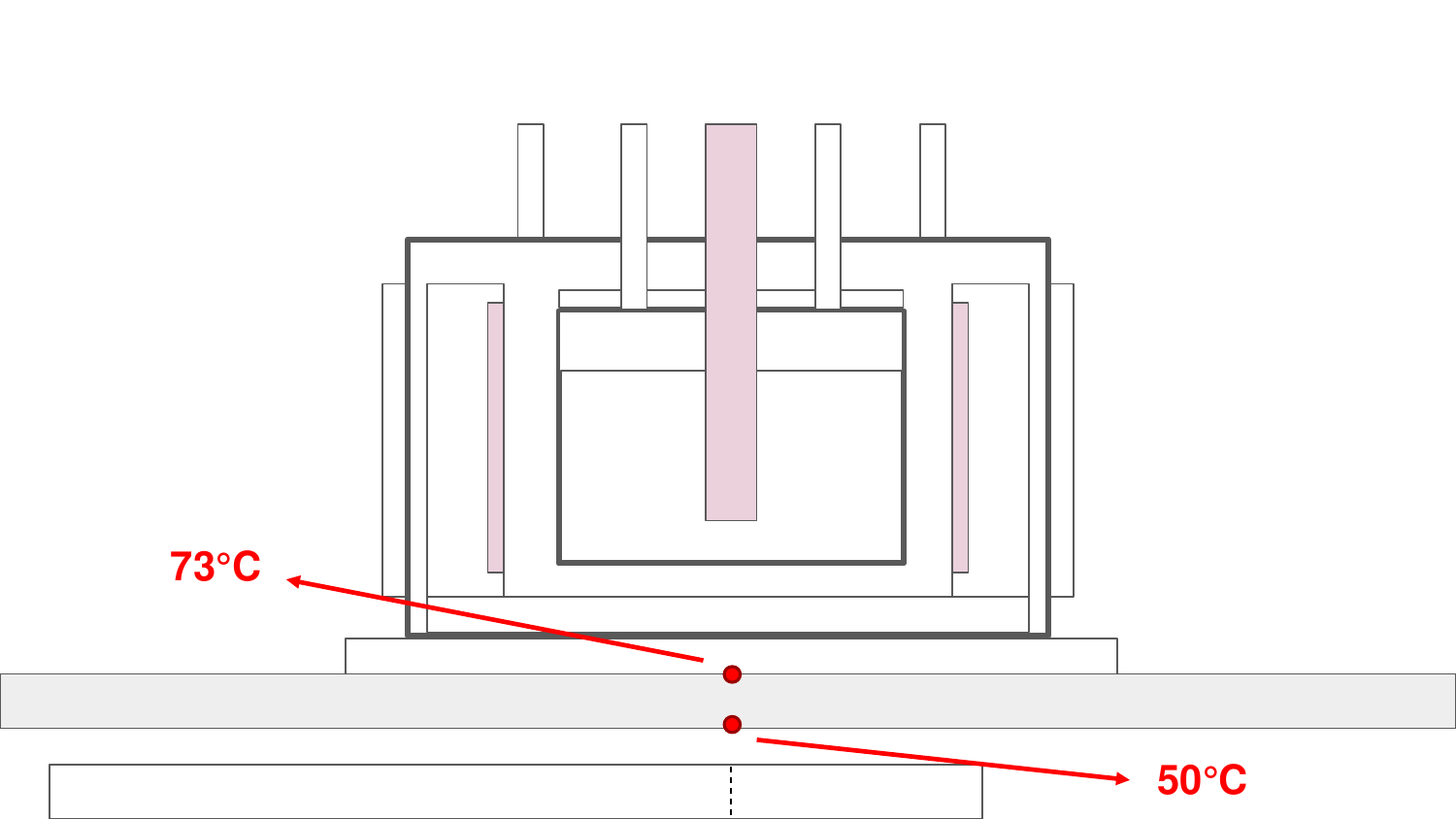}
         \caption{Approximate typical steady-state axial distribution of temperature under the vessel assembly.}
         \label{fig:axial_temperature_distribution}
     \end{subfigure}
    \caption{Cross-section diagram of the vessel assembly showing system components and the temperature distribution, not to scale.}
    \label{fig:vessel_xsection}
\end{figure}

The temperature distribution in the system was recorded using several K-type thermocouples:
\begin{itemize}
    \item Embedded in the cartridge heater;
    \item Probe embedded in the furnace, midway through its height;
    \item Inconel sheathed 1/8" (\SI{0.3175}{cm}) probe inserted vertically into a re-entrant tube in the crucible, 2.25" (\SI{5.715}{cm}) away from the vessel axis, reaching midway through the salt volume; 
    \item Inconel sheathed 1/8" (\SI{0.3175}{cm}) probe inserted vertically into a conduit welded to the outside of the crucible, reaching its bottom;
    \item Surface probe attached to the outside wall of the outer vessel, under the silica jacked, midway through its height;
    \item Surface probe attached to the outside of the silica jacket;
    \item Surface probe attached to the top of the epoxy surface, under the kaowool insulation, on vessel axis;
    \item Surface probe attached to the underside of the epoxy, on vessel axis.
\end{itemize}
The typical steady-state distribution of temperature in the system is shown on Figure \ref{fig:vessel_xsection}. \par

In initial testing of the system performed with the cartridge heater alone, and with a 1" (\SI{2.54}{cm}) thick layer of kaowool lining the inside of the outer vessel instead of the furnace, it was not possible to reach the melting temperature across the entire volume of the salt without exceeding the temperature rating of the heater.
With the addition of the furnace a minimum temperature of 630$^\circ$C was achieved throughout the salt, which is above the 510$^\circ$C melting temperature of eutectic ClLiF, but still under the targeted salt temperature of 700$^\circ$C. This target was set for consistency with the prior \SI{100}{mL} BABY experiments \cite{delaporte-mathurin_advancing_2025}, as well as because it falls within the typically proposed operating temperature range for a LIB blanket \cite{sorbom_arc_2015}.
The inability of the furnace to reach the setpoint at full heating power and the relatively high temperature of the outer vessel wall suggest that heat is dominantly lost radially. \par

To maintain the temperature around the generator within its operating envelope, a blower and a fan were installed under the workbench to provide airflow for cooling.
Due to the unknown composition and glass transition temperature of the epoxy plate an alarm PID controller was installed to the heating power supply, which would cut power to the cartridge heater if the temperature at the top of the epoxy reaches 75$^\circ$C.
This value was selected as the lowest temperature which was typically not reached during normal operation, while not causing any apparent damage to the workbench. \par

\subsubsection{Gas handling system}

The majority of the gas system was composed of 1/4" (\SI{0.635}{cm}) OD 316 SS tubing assembled using Swagelok compression fittings, adapted to 1/2" (\SI{1.27}{cm}) connections at the vessel assembly.
An inlet manifold supplied gas at a working pressure of 10 PSIG (\SI{68.9}{kPa} gauge), and conditioned it using Restek oxygen and moisture scrubber columns.
The flow rates in the IV and OV streams were maintained using Alicat mass flow controllers in the range of 0-50 SCCM.
In all presented runs the IV flow rate was set at 30 SCCM, and the OV at 50 SCCM.
Two gas compositions were available, UHP He or 1000 ppm H\textsubscript{2} bal. He, and the same selected composition was supplied to both streams. \par

To enable sparging of the IV stream gas through the salt, a 1/8" (\SI{0.3175}{cm}) OD, 0.055" (\SI{0.1397}{cm}) ID Inconel-600 tube was fitted into the 1/2" (\SI{1.27}{cm}) crucible inlet tube through a Swagelok Ultra-Torr fitting.
Once the salt was molten, the fitting was loosened sightly and the sparger tube lowered into the salt, down to around 1/2" (\SI{1.27}{cm}) above the bottom of the crucible.
A system of valves enabled routing the gas either through the sparger or above the salt as cover gas. \par

Downstream from the vessel, between it and the bubblers, a series of progressively coarser 316 SS sintered element inline particulate filters from Swagelok was placed.
In the IV stream, the series consisted of \SI{15}{\micro \metre}, \SI{7}{\micro\metre}, and \SI{2}{\micro\metre} nominal pore size filters, while the OV stream only contained a \SI{15}{\micro\metre} filter. \par

The oxidation furnaces between the bubblers contain Pd-on-alumina beds at \SI{450}{\celsius}, which require the addition of oxygen to the gas stream in order to convert molecular tritium species to tritiated water.
To facilitate this, 10 SCCM of air was mixed into each of the gas streams before passing through the bubblers.
Proper mixing was verified using mass flow meters that measured the combined sample gas and air flow rates. \par

Analog pressure gauges were distributed at several positions along the system, including directly at the inlet to the sparger/crucible and at the lid of the outer vessel.
Typical operating pressure was around 3 PSIG (\SI{20.7}{kPa} gauge) in both streams.
During bubbler sampling, the sample gas streams were diverted from the bubblers and into the atmosphere to reduce water migration and splashing.
To stabilize the temporary disturbances to the system pressure and flow rate this procedure caused, needle valves were added at the outlets of the diversion branches. \par

The temperature of tubing downstream from the vessel decreased rapidly, indicating that the gas cools down shortly after leaving the hot section of the system.
No apparent clogging or constrictions were observed due to salt vapor condensation or particulate migration, either at the filters or between them and the crucible. \par

Since the furnace was positioned within the OV gas stream, off-gassing from the bare heating element or the insulation layer was observed during the first several days of operation, which contaminated the bubbler water, discolouring it and changing its pH.
This effect dissipated before the first reported run. \par  

\begin{figure*}[p]
    \centering
    \includegraphics[width=\textwidth,height=0.95\textheight,keepaspectratio]{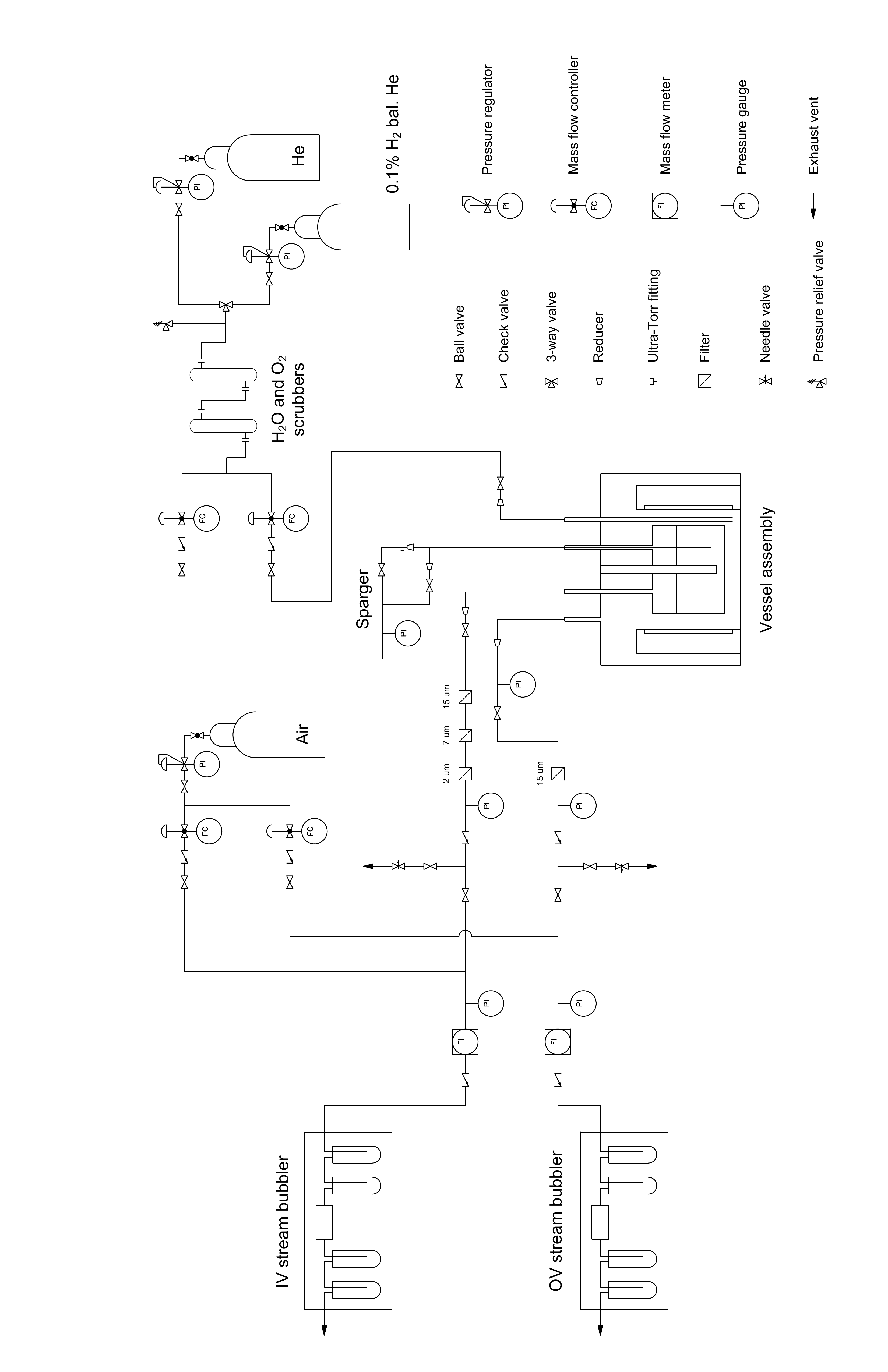}
    \caption{P\&ID diagram of the 1L BABY gas system.}
    \label{fig:pid_diagram}
\end{figure*}

The full P\&ID diagram of the 1 L BABY gas system is shown on Figure \ref{fig:pid_diagram}.

\subsubsection{Tritium detection}

The tritium detection technique was identical to the BABY \SI{100}{mL} experiment \cite{delaporte-mathurin_advancing_2025}.
The gas streams containing tritium were passed through water bubblers (one per stream) to collect HTO, TF, TCl, HT, T2.
The water was then mixed with the liquid scintillation cocktail, and the tritium activity was counted on a liquid scintillation counter. \par

A typical sample in the runs reported here consisted of 10 mL of the aquaeous component and 10 mL of PerkinElmer UltimaGold LLT scintillation cocktail, and was counted on a PerkinElmer TriCarb 5110TR scintillation counter for 100 minutes.
With a background of around 4 CPM per sample and at a counting efficiency of around 25\% in the 0.5-5.5 keV energy range, the minimum detectable activity (MDA) in a given sample was under 0.1 Bq, according to the MDA calculation from \cite{lannunziata_chapter_2012}.

\subsubsection{Neutron detection: activation foils}

Activation foil analysis is a method for detecting neutrons by triggering nuclear reactions in specially chosen target materials, commonly referred to as activation foils \cite{lee_determination_2019, greenberg_neutron_2011}.

Activation foil analysis was used to quantify the total neutron fluence during each run and served as the reference for the denominator in the measured TBR calculation, following the same methodology as outlined in the 100~mL BABY campaign \cite{delaporte-mathurin_advancing_2025}.
Niobium (Nb) and zirconium (Zr) foils were selected as activation materials due to their well-characterised neutron-induced reactions with high energy thresholds.
The primary reaction measured from neutron activation of the Nb foils was the ${}^{93}$Nb(n,2n) reaction, which has a threshold of 8.93 MeV, resulting in the production of ${}^{92m}$Nb, which beta decays with a half-life of 10.15 days and emits a 934 keV gamma ray.
The reaction measured in the Zr foils was the ${}^{90}$Zr(n,2n) reaction, which has an even higher threshold of 12.1 MeV and produces ${}^{89}$Zr, which beta decays with a half-life of 78.4 hours, emitting a 909 keV gamma ray \cite{noauthor_livechart_nodate}. 

The foils were placed approximately \SI{5}{cm} above the neutron emission point, directly on top and below the sealed neutron source tube.
During irradiation, neutron interactions produce metastable isotopes such as \ce{^{92m}Nb}, whose population increases over the course of the run and then decays exponentially after the neutron source is turned off.

\begin{figure}[h]
    \centering
    \includegraphics[width=1\linewidth]{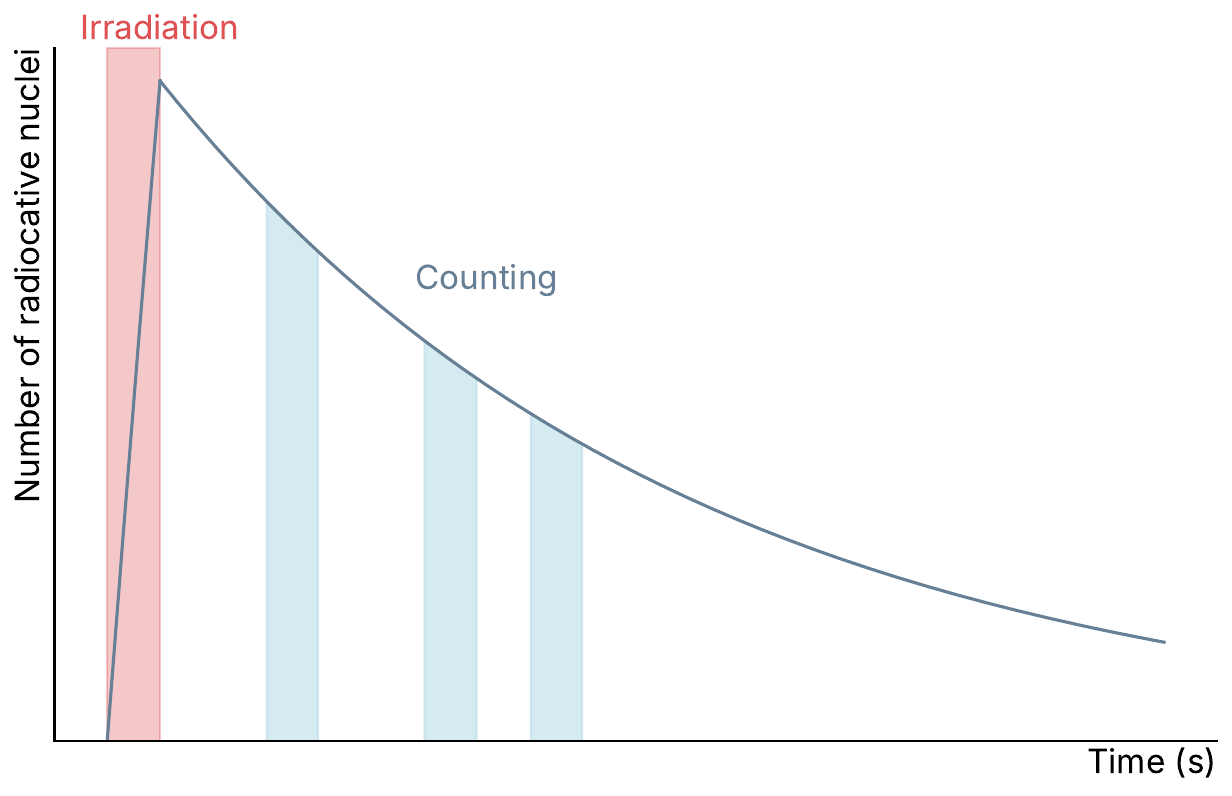}
    \caption{Schematic representation of the number of radioactive nuclei (e.g., \ce{^{92m}Nb}) in the foil over time. The build-up during neutron irradiation is followed by exponential decay. The shaded regions indicate the periods over which decays are counted.}
    \label{fig:irradiation-counting}
\end{figure}

\Cref{fig:irradiation-counting} illustrates the typical time evolution of the radioactive isotope population, including the irradiation phase, decay period, and the counting windows.
The shaded area represents the portion of \ce{^{92m}Nb} decays measured using sodium iodide (NaI) scintillators.
Each foil was analysed between one to three times using two 4" x 4" x 4" (\SI{10.16}{cm} x \SI{10.16}{cm} x \SI{10.16}{cm}) NaI detectors surrounded by lead bricks, with energy and efficiency calibration performed using dedicated check sources to ensure reliable identification of characteristic gamma lines and construct a detector efficiency curve each time the NaI detectors were used.

\begin{equation}
    N \approx C \ \sigma \ \varphi
\end{equation}

\begin{align}
    N &= A \ f_{\text{time}} \ f_{\text{self}} \ f_{\text{spec}} \ \sigma \varphi \\
    &= A \ \left[ \frac{(1 - e^{-\lambda t_{\text{irr}}}) \ e^{-\lambda t_d} \ (1 - e^{-\lambda t_c})}{\lambda} \ \left( \frac{t_{\text{live}}}{t_c} \right) \right] \notag \\
    &\quad \ \left[ \frac{1 - e^{-\mu(E) \rho x}}{\mu(E) \rho x} \right] \ \left[ \varepsilon(E) \ \mathrm{BR}(E) \right] \  \sigma \varphi
\end{align}

\begin{itemize}
    \item $N$ – Net count of a gamma-ray peak [counts]
    \item $C$ – Calibration constant relating neutron fluence to counts
    \item $\sigma$ – Neutron activation cross section [\si{\centi\metre\squared}]
    \item $\varphi$ – Neutron flux [\si{n\per\centi\metre\squared\per\second}]
    \item $A$ – Number of activated nuclei in the foil [\#]
    \item $\lambda$ – Decay constant of the radioactive isotope [\si{\per\second}]
    \item $t_{\text{irr}}$ – Irradiation time [\si{\second}]
    \item $t_d$ – Decay time between end of irradiation and start of counting [\si{\second}]
    \item $t_c$ – Real (clock) counting time [\si{\second}]
    \item $t_{\text{live}}$ – Live counting time (corrected for dead time) [\si{\second}]
    \item $\mu(E)$ – Mass attenuation coefficient at photon energy $E$ [\si{\centi\metre\squared\per\gram}]
    \item $\rho$ – Density of the activation foil [\si{\gram\per\centi\metre\cubed}]
    \item $x$ – Thickness of the activation foil [\si{\centi\metre}]
    \item $\varepsilon(E)$ – Absolute detection efficiency of the NaI detector at energy $E$ [unitless]
    \item $BR(E)$ – Branching ratio for gamma emission at energy $E$ [unitless]
    \item $f_{\text{time}}$ – Time correction factor (accounting for decay and counting periods) [unitless]
    \item $f_{\text{self}}$ – Self-absorption correction factor in the foil [unitless]
    \item $f_{\text{spec}}$ – Spectral correction factor (due to neutron energy distribution) [unitless]
\end{itemize}

For fluence estimation, an isotropic neutron source assumption was applied.

Amongst the sources of uncertainty contributing to the measured TBR, activation foil analysis currently dominates.
This is due to uncertainties in foil placement, detection efficiency, and cross-section data. 
Ongoing efforts aim to reduce this uncertainty through improved counting statistics and refined calibration procedures.



\subsubsection{Neutron detection: Proton Recoil Telescope}

The proton recoil telescope (PRT) is a neutron spectrometer \cite{weiss_fusion_2024} that is used in the BABY experiment.
Detection of neutrons using diamonds requires that the neutrons react with the carbon atoms in the diamond lattice to produce alpha particles.
This reaction has a 6.92 MeV energy threshold, so any signal that is then produced by the alpha particle’s creation of e-h pairs corresponds to an energy that is at least 6.92 MeV less than the energy of the original neutron.

In order to capture the full energy of an incident neutron, a converter foil is introduced.
A thin film of polyethylene plastic is placed in front of the diamond detection volume.
The plastic material is chosen so that it will undergo an elastic (n,p) reaction, transferring up to the full neutron energy to a proton that is scattered into the detector volume.
Then, the proton can deposit all of its energy by interacting with the electrons in the diamond. Furthermore, the detection volume is broken into four separate diamonds aligned in a telescope configuration.
By doing this, coincident signals that occur in multiple diamonds can be trusted to have been caused by relevant particles, i.e. the protons that were scattered into the telescope with approximately the full energy of the original neutron.
This allows for the filtration of background signals that are caused by neutron elastic scattering in the diamond, as those signals would likely be constrained to a single diamond volume.

\subsection{Modelling}
\subsubsection{Neutronics modelling}

\begin{figure}[h!]
     \centering
     \begin{subfigure}[b]{0.8\linewidth}
         \centering
         \includegraphics[width=\linewidth]{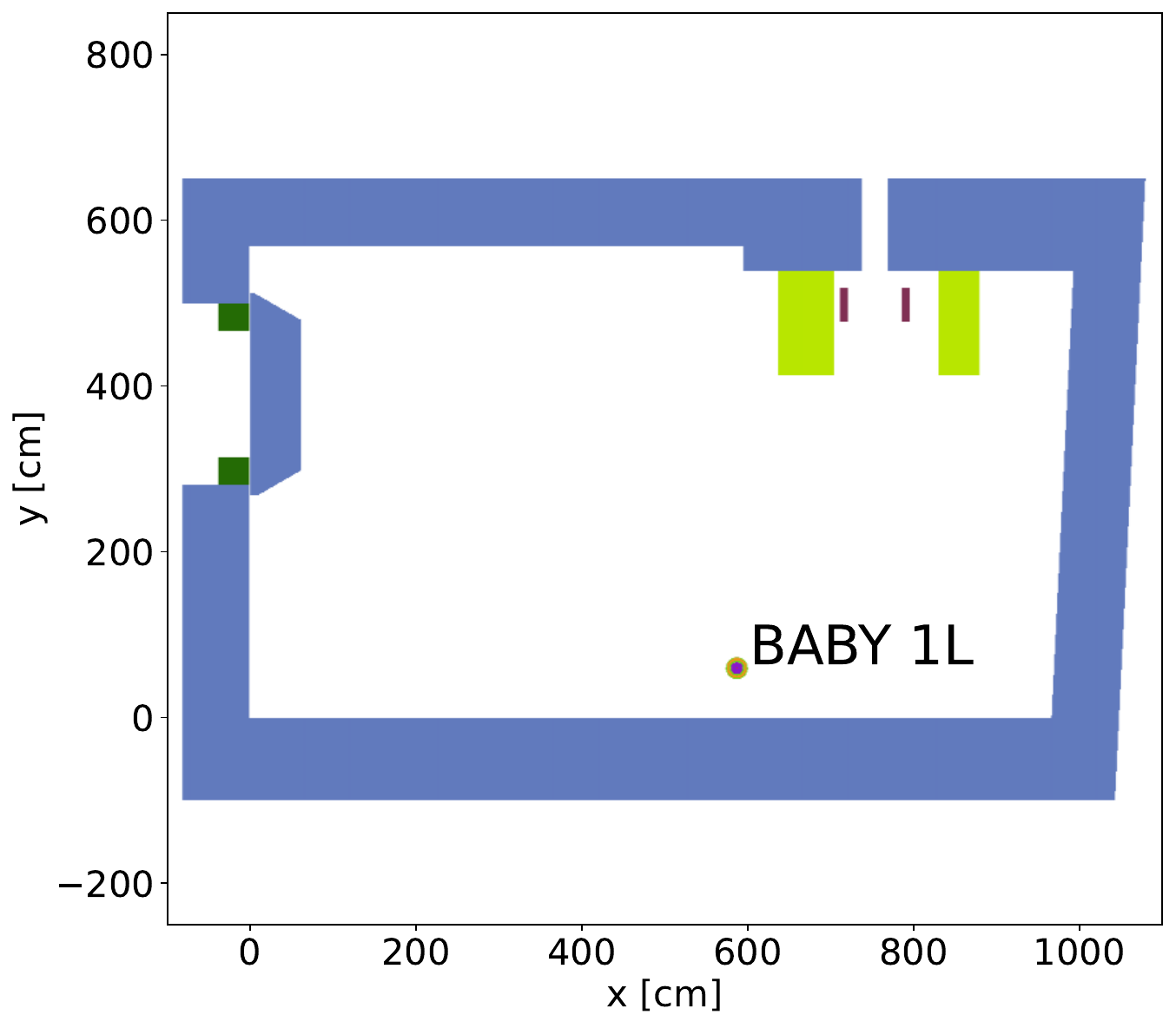}
         \caption{Top view of the vault laboratory model}
     \end{subfigure}
     \begin{subfigure}[b]{0.8\linewidth}
         \centering
         \includegraphics[width=\linewidth]{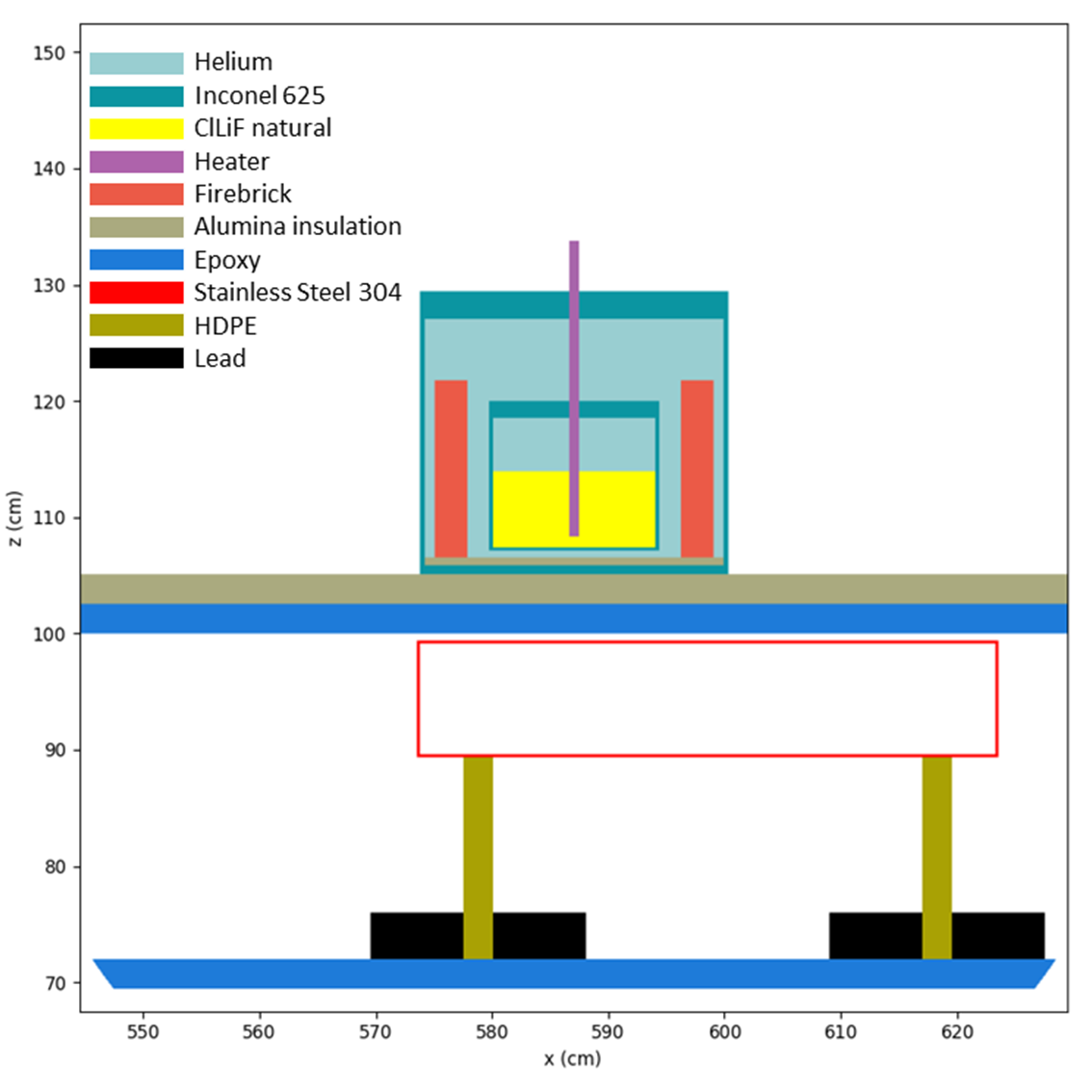}
         \caption{Close view of the 1 L BABY experiment model including neutron source, lead blocks, vessels, furnace and tables ($y=\SI{60}{cm}$).}
     \end{subfigure}
    \caption{Geometry of the OpenMC neutronics model.}
    \label{fig:neutron-geometry}
\end{figure}

The neutronics model of the BABY 1L experimental setup is developed using the OpenMC Monte Carlo code \cite{ROMANO201590} and the ENDF/B-VIII.0 evaluated nuclear data library \cite{conlin2018release}.

The geometry is constructed using a constructive solid geometry (CSG) approach, which enables detailed control over components and spatial regions.
A high-fidelity representation of the experimental assembly is included, featuring internal structures, shielding elements, and support materials.
The neutron source is positioned at the base of the device and mechanically supported by lead bricks and high-density polyethylene (HDPE) blocks.
The setup rests on a support table, modeled along with thermal insulation layers.
All geometrical features are based on realistic dimensions and material properties derived from experimental design data and reference compendia \cite{mcconn2011compendium}.
Compared to the \SI{100}{mL} BABY model \cite{delaporte2025advancing}, the full MIT Vault Laboratory is also included in the model to better capture the neutron backscattering effects caused by the surrounding walls.
The layout and materials of the room strongly influence this contribution.
Movable elements within the Vault, such as other experiments or furniture, are excluded due to their variable configuration.

The resulting geometry is shown in \Cref{fig:neutron-geometry}.
Materials definitions and input files are openly available in the LIBRA project GitHub repository \url{github.com/LIBRA-project}.

\subsubsection{Tritium release modelling}
\paragraph{0D model}

The tritium release can be modelled by a simple ODE:

\begin{equation}
    V \frac{d c_\mathrm{salt}}{dt} = S - Q_\mathrm{inner \ vessel} - Q_\mathrm{outer \ vessel} \label{eq:tritium ode}
\end{equation}
where $V$ is the salt volume and $S$ is the source of tritium in \si{T\per\second} expressed by:

\begin{equation}
    S = \mathrm{TBR} \cdot \Gamma_\mathrm{n} 
\end{equation}
where $\Gamma_\mathrm{n} $ is the neutron rate in \si{neutron\per\second}.

The tritium release rates $Q_i$ in \si{T\per\second} are expressed as:
\begin{align}
    Q_i &= A_i \ k_i \ (c_\mathrm{salt} - c_\mathrm{external}) \\
    &\approx A_i \ k_i \ c_\mathrm{salt}
\end{align}
where $A_i$ is the surface area of each release pathway in \si{\metre\squared}, $c_\mathrm{external}$ is the tritium concentration in the gas phase (assumed negligible), and $k_i$ is the mass transport coefficient in \si{\metre\per\second}.

For each experiment, the mass transport coefficients $k_i$ and $\mathrm{TBR}$ are set as free parameters to fit the measured tritium release.






\subsection{Analysis workflow}


To streamline and standardise the analysis of complex - and growing - experimental datasets in the BABY campaign, we developed an integrated software suite named \texttt{libra-toolbox}.
This open-source toolkit provides a unified framework for handling each component of the experimental workflow, ranging from data acquisition to physics-based modelling.
The workflow begins with experimental measurements of total tritium production and neutron yield, from which the measured TBR is derived.
These values are compared against predictions from neutron transport simulations using OpenMC, which incorporate detailed experimental geometries.

\texttt{libra-toolbox} supports automated generation and post-processing of OpenMC neutronics models, including tally extraction and normalisation to experimental neutron sources. 
It also includes modules for low-level analysis of tritium release data from liquid scintillation counting (LSC), activation foil analysis, and real-time neutron detection using the PRT, including coincidence analysis for energy discrimination.
Moreover, the toolbox interfaces with tritium transport simulations, both in simplified 0D models and in spatially resolved simulations using FESTIM.
Mass transport coefficients extracted from these models can be compared directly with experimental release curves to validate transport assumptions.

This modular, reproducible workflow not only accelerates data interpretation but also enables consistent comparison between runs and across campaigns, facilitating deeper insights into tritium behaviour and breeding performance under varying conditions.

The LIBRA framework also supports transparency, reproducibility, and modularity.
Each experiment is hosted in a dedicated, version-controlled repository that includes simulation, analysis, and documentation.
All workflows are automated and validated through continuous integration.
Data and metadata comply with findable, accessible, interoperable, and reusable (FAIR) principles, ensuring full traceability and reusability.
Documentation is provided through embedded Jupyter Notebooks, allowing users to reproduce results with minimal setup.

\begin{figure*}[h!]
  \centering
  \begin{subfigure}{0.8\textwidth}
    \includegraphics[width=\textwidth]{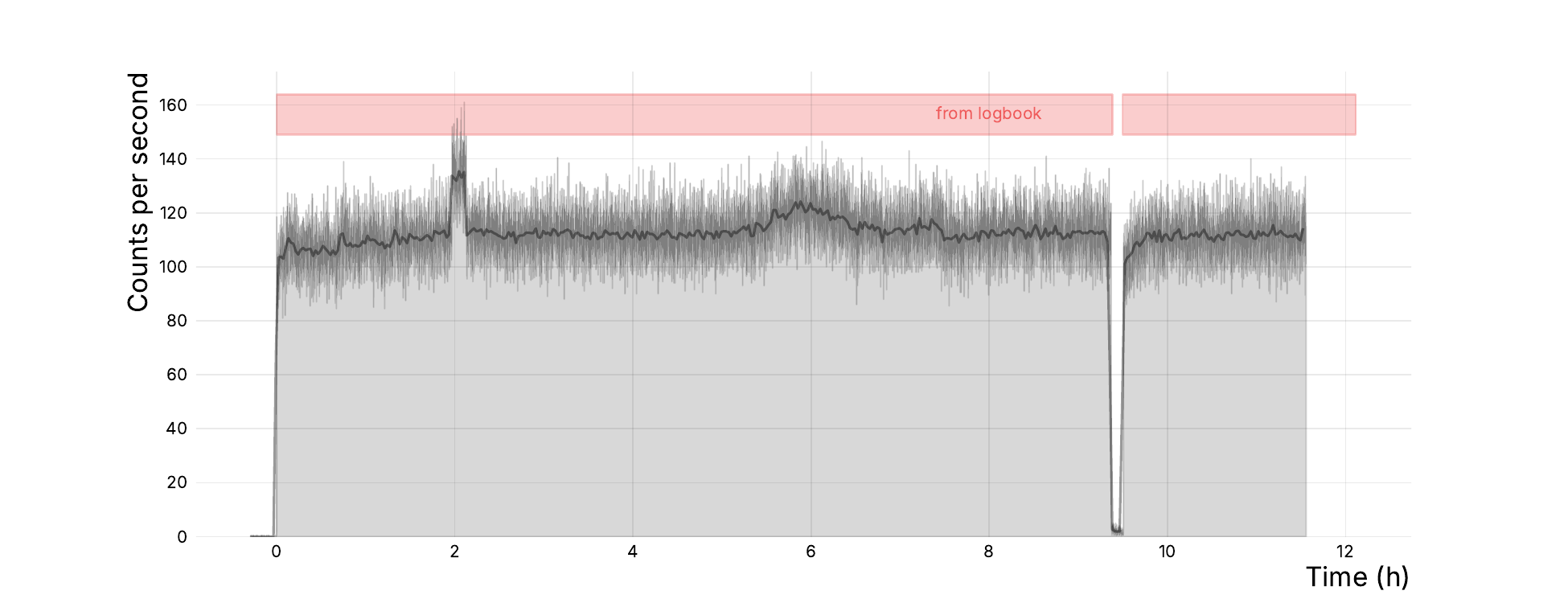}
    \caption{Temporal evolution of the \SI{14}{MeV} neutron count rate.}
    \label{fig:sub-a}
  \end{subfigure}
  \hfill
  \begin{subfigure}{0.45\textwidth}
    \includegraphics[width=\textwidth]{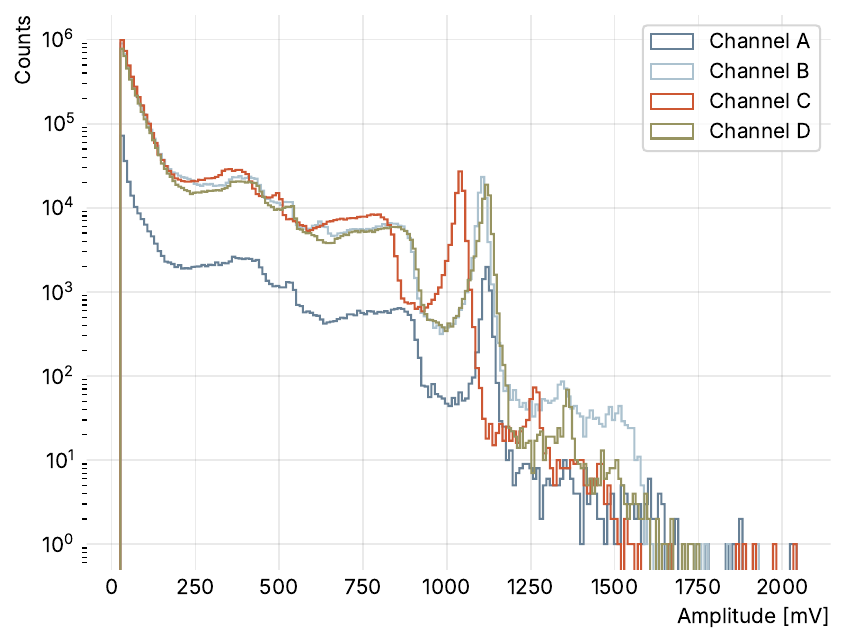}
    \caption{Uncalibrated particle energy spectra in the four diamonds (before coincidence calculations).}
    \label{fig:sub-b}
  \end{subfigure}
  \hfill
  \begin{subfigure}{0.45\textwidth}
    \includegraphics[width=\textwidth]{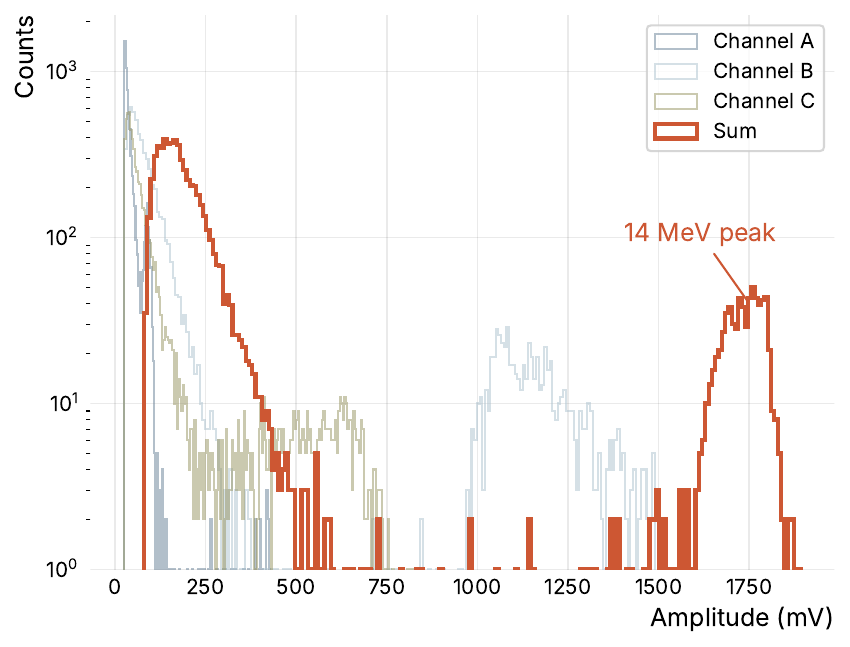}
    \caption{Uncalibrated particle energy spectra in the four diamonds (after coincidence calculations).}
    \label{fig:sub-c}
  \end{subfigure}
  \caption{Proton Recoil Diamond Telescope results during BABY 1L run 3.}
  \label{fig:PRT results run 3}
\end{figure*}

\section{Results}\label{results}

This paper will focus on the data generated from BABY 1L runs 1 to 4.
Below are the key differences between each run:

\begin{itemize}
    \item Run 1: Pure He. Data: \cite{remi_delaporte_mathurin_2025_17160928}
    \item Run 2: Pure He + higher neutron fluence. Data: \cite{remi_delaporte_mathurin_2025_17160748}
    \item Run 3: Pure He + sparging; switch to 1000 ppm \ce{H_2} mid-run. Data: \cite{remi_delaporte_mathurin_2025_17160749}
    \item Run 4: 1000 ppm \ce{H_2} (from the start) + sparging. Data: \cite{remi_delaporte_mathurin_2025_17160762}
\end{itemize}

\subsection{Neutron measurements}

The use of a PRT in the BABY 1L campaign enabled real-time monitoring of the neutron emission rate throughout each run (see Figure \ref{fig:PRT results run 3}).
This continuous measurement capability allowed us to observe fine temporal variations in the neutron flux that would not have been captured using traditional logging methods, such as simply recording the start and end times of generator operation.
Additionally, the coincidence counting feature of the PRT provides spectral information, confirming that the emitted neutron spectrum is dominated by monoenergetic \SI{14}{MeV} neutrons characteristic of DT fusion reactions.
This confirmation of spectral purity is essential for accurate interpretation of activation and tritium production data.

\subsection{TBR measurements}

Neutron measurements were performed for all four runs of the BABY 1L experiment, and the corresponding total tritium collected is summarised in \cref{tab:tbr}. From these measurements, the Tritium Breeding Ratio (TBR) was computed for each individual run as well as for the aggregate data.

The average measured TBR for the 1L configuration across all runs was \num{2.46e-3} — a sixfold increase compared to the \num{3.61e-4} value obtained in the earlier 100~mL experiments \cite{delaporte-mathurin_advancing_2025}. This significant enhancement can be largely attributed to the greater solid angle coverage of the 1L volume, which increases the likelihood of neutron interactions.

\begin{figure*}[h]
    \centering
    \includegraphics[width=0.7\linewidth]{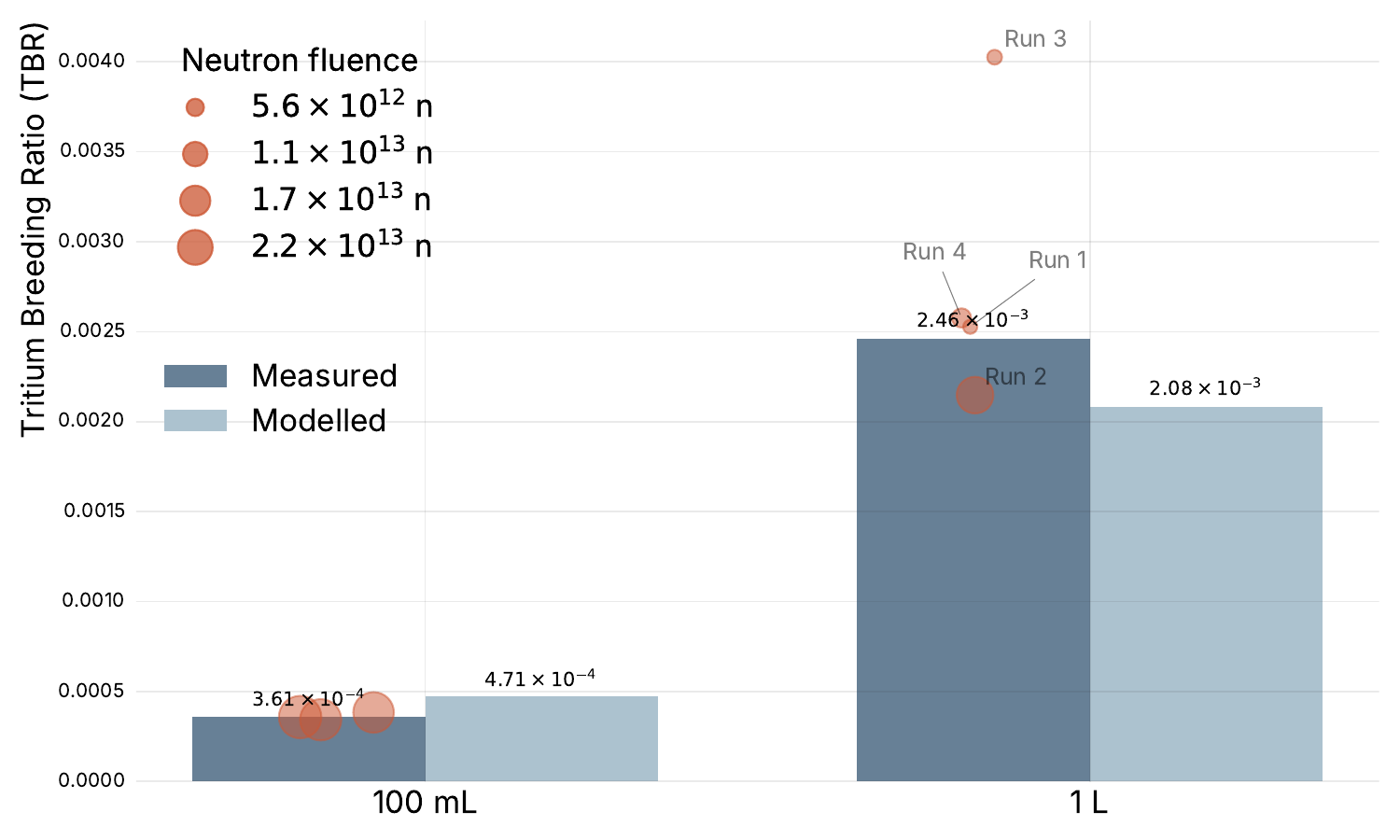}
    \caption{Comparison of the modelled TBR (from OpenMC) and experimental measurements. Bars for measured TBR correspond to the average over all runs weighted by neutron fluence. Red dots represent individual experiments.}
    \label{fig:tbr-comparison-with-openmc}
\end{figure*}

A comparison between the experimentally derived TBR values and those predicted by OpenMC simulations shows very good agreement (see \cref{fig:tbr-comparison-with-openmc}).
The modelled TBR for the 1 L case was \num{2.08e-3}, closely matching the experimental measurements.
While individual runs exhibit some variability, all measured TBR values fall within reasonable proximity to simulation results, supporting the accuracy of both the experimental method and the computational model.

Additional details, including neutron fluence and tritium production for each run, are provided in \cref{tab:tbr}.
The high value for the TBR measured in Run 3 is the result of a high tritium release.
This will be explained in more detail in \cref{influence_h2}.

\begin{table*}
    \caption{Tritium Breeding Ratio (TBR) for different runs and volumes.}
    \label{tab:tbr}
    \centering
    \begin{tabular}{lrrrr}
         & Measured TBR & Modelled TBR & neutron fluence & tritium production \\
        \midrule
        100 mL \cite{delaporte-mathurin_advancing_2025} & $3.61 \times 10^{-4}$ & $4.71 \times 10^{-4}$ &  & \\
        1 L (all)\footnotemark & $2.45 \times 10^{-3}$ & $2.8 \times 10^{-3}$ & $4.00 \times 10^{13}$ & $9.78 \times 10^{10}$ \\
        1L run 1 & $2.39 \times 10^{-3}$ & $2.08 \times 10^{-3}$ & $3.95 \times 10^{12}$ & $9.45 \times 10^{9}$ \\
        1L run 2 & $2.15 \times 10^{-3}$ & $2.08 \times 10^{-3}$ & $2.50 \times 10^{13}$ & $5.36 \times 10^{10}$ \\
        1L run 3 & $4.03 \times 10^{-3}$ & $2.08 \times 10^{-3}$ & $4.12 \times 10^{12}$ & $1.66 \times 10^{10}$ \\
        1L run 4 & $2.58 \times 10^{-3}$ & $2.08 \times 10^{-3}$ & $6.91 \times 10^{12}$ & $1.81 \times 10^{10}$ \\
    \end{tabular}
\end{table*}
\footnotetext{Average over all runs weighted by neutron fluence.}

\subsection{Release dynamics}\label{dynamics}

\begin{figure}[h]
    \centering
    \includegraphics[width=1\linewidth]{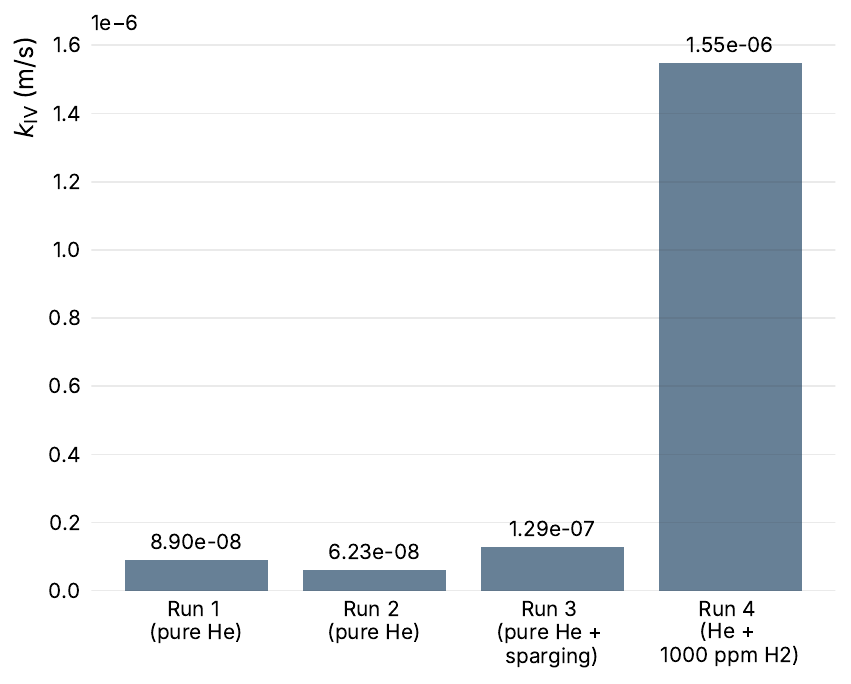}
    \caption{Comparison of the IV mass transfer coefficient $k_\mathrm{IV}$ between runs 1-4.}
    \label{fig:k_IV}
\end{figure}

Compared to similar conditions (pure He) in previous experiments using a \SI{100}{mL} salt volume, the release of tritium in the BABY 1L runs was noticeably slower.
This is consistent with expectations for larger systems where the diffusion/transport path is longer, thereby slowing mass transfer to the gas phase.

Interestingly, among the 1L runs, Run~2 exhibited a markedly slower release profile than Runs~1 and~3. A possible explanation for this anomaly is an operational incident during Run~2: the heater malfunctioned, resulting in the salt freezing mid-run. The salt was later re-melted, which may have affected the dynamics of tritium transport in that particular trial.

Additionally, we explored the impact of helium sparging as a potential enhancement mechanism for tritium removal. During the first 19 days of Run~3, pure helium was bubbled through the salt. However, we did not observe a significant increase in tritium release rate. This limited effect may be due to the simplicity of the sparging setup, which likely did not induce sufficient agitation or circulation in the molten salt to substantially enhance convective mass transfer.

Overall, the release dynamics measured across the runs reinforce the diffusion-limited nature of tritium transport in molten salts, and suggest that significant enhancements to removal rates would require more effective agitation or alternative engineered solutions.

To better understand the mass transfer mechanisms at play, we consider the Sherwood number, defined as:
\begin{equation}
    \mathrm{Sh} = \frac{k}{D/L}
    \label{eq:sherwood}
\end{equation}
where $k$ is the mass transfer coefficient (\si{\metre\per\second}), $D$ is the diffusion coefficient (\si{\metre\squared\per\second}), and $L$ is the characteristic length scale (\si{\metre}).

\begin{figure}
    \centering
    \includegraphics[width=1\linewidth]{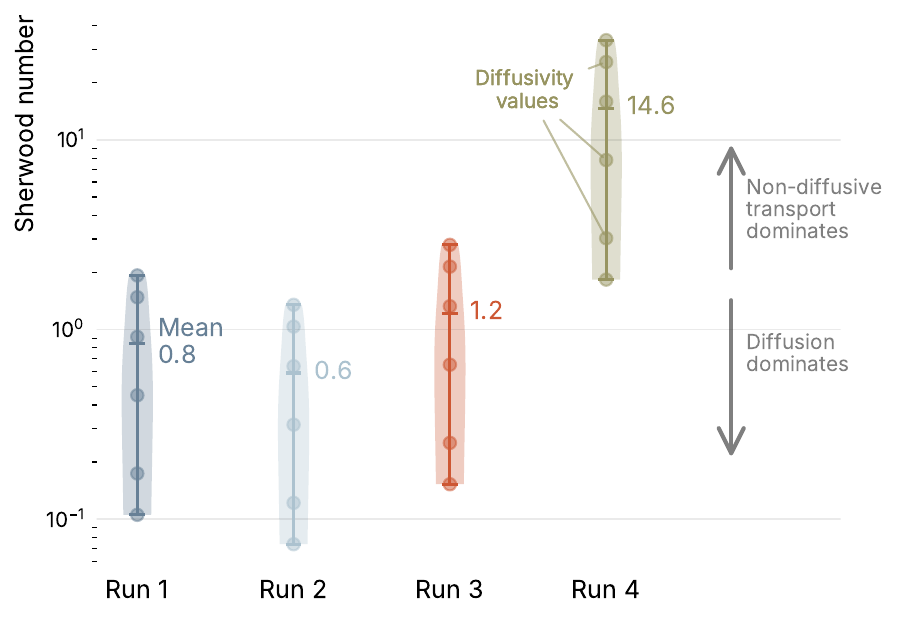}
    \caption{Comparison of the Sherwood number for each run. Each point correspond to a diffusivity value in the HTM database \cite{}.}
    \label{fig:sherwood}
\end{figure}

The Sherwood number quantifies the relative importance of convective versus diffusive transport. In this context, $\mathrm{Sh} \ll 1$ would indicate a diffusion-limited regime.
Previous work by Fukada et al.~\cite{fukada_initial_2002} has already shown that tritium transport in molten salts (in their case FLiBe) can be diffusion-limited, and our observations are consistent with this finding.
Runs 1, 2, and 3 exhibited $\mathrm{Sh} \approx 1$ whereas Run 4 showed $\mathrm{Sh} > 1$ (see \cref{fig:sherwood}).
The explanation for this different regime will be given in the next section.

\subsection{Influence of \ce{H2} in the sweep and cover gas}\label{influence_h2}

\begin{figure*}[h]
    \centering
    \begin{minipage}[b]{0.4\textwidth}
        \centering
        \includegraphics[width=\linewidth]{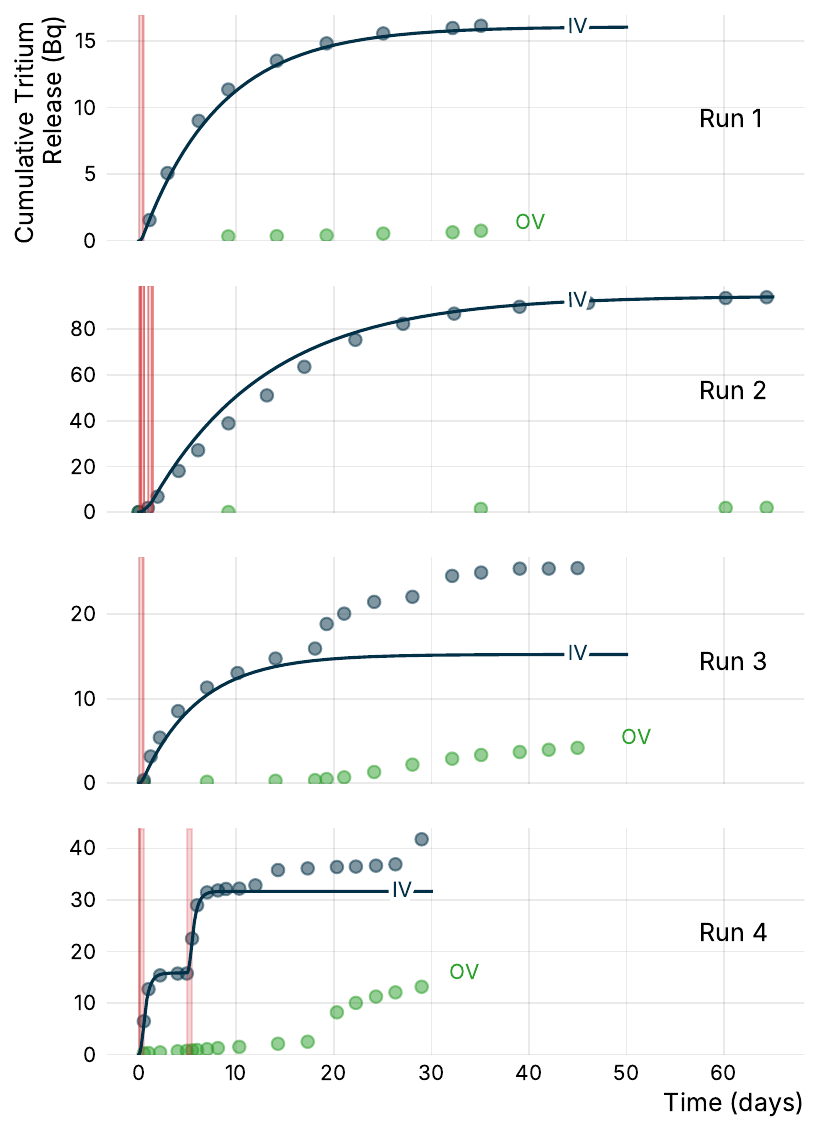}
        \subcaption{Comparison of tritium release dynamics for Runs 1 to 4.}
        \label{fig:release-runs-1-4}
    \end{minipage}
    \hspace{5em}
    \begin{minipage}[b]{0.40\textwidth}
        \centering
        \begin{subfigure}[t]{\linewidth}
            \centering
            \includegraphics[width=\linewidth]{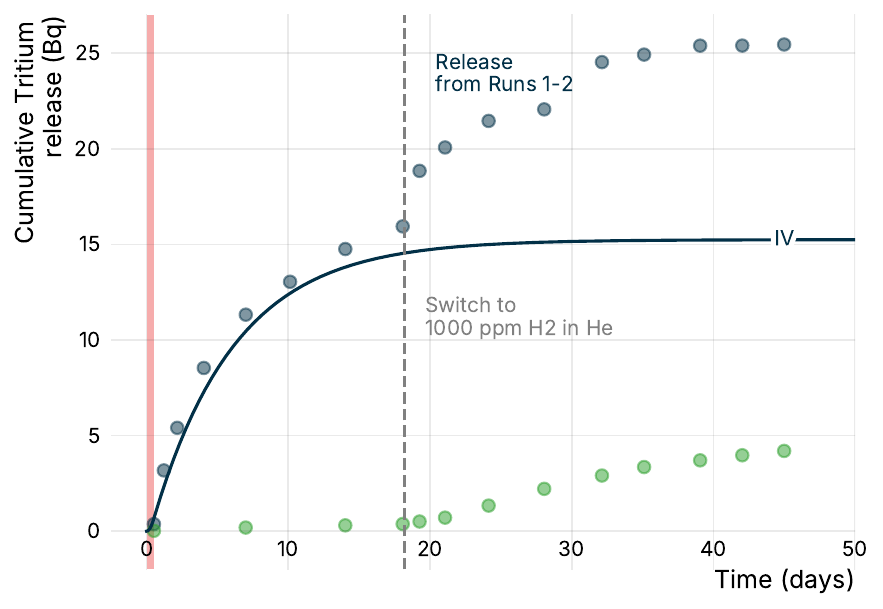}
            \caption{Run 3.}
            \label{fig:release-run-3}
        \end{subfigure}
        \vfill
        \begin{subfigure}[b]{\linewidth}
            \centering
            \includegraphics[width=\linewidth]{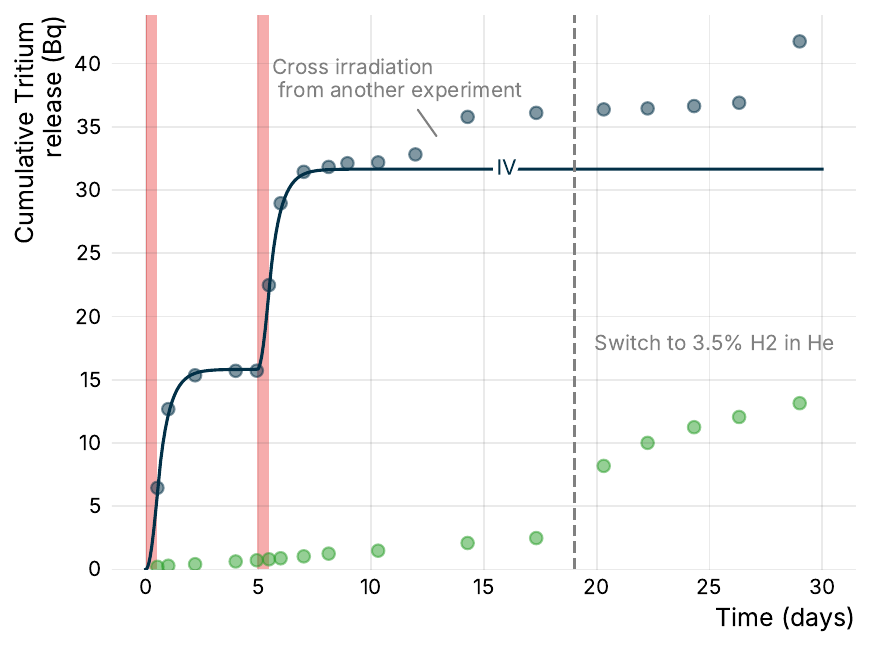}
            \caption{Run 4.}
            \label{fig:release-run-4}
        \end{subfigure}
    \end{minipage}

    \caption{Overview of tritium release behaviour during the BABY 1L runs. The solid line corresponds to the model prediction for the IV release.}
    \label{fig:composite-release}
\end{figure*}

Runs 3 and 4 of the BABY 1L campaign explored the influence of hydrogen on tritium release dynamics.
In Run 3, the experiment began with pure helium as the carrier gas.
Midway through the run, hydrogen (1000 ppm \ce{H2}) was introduced, resulting in a noticeable change in the release behaviour (see \cref{fig:release-run-3}).
This shift appeared to activate an additional release pathway associated with the outer vessel.
We hypothesise that this behaviour is due to isotopic exchange between the introduced \ce{H2} and tritium adsorbed on surfaces (e.g. outer surface of the crucible, surface of cold tubing, etc.), forming \ce{HT}:

\begin{equation}
    \ce{H2 + T (ad) <-> HT + H (ad)}
\end{equation}

Further support for this hypothesis was obtained in Run 4, where increasing the \ce{H2} concentration from \SI{1000}{ppm} to \SI{3.5}{\percent} on day 19 led to an enhanced release from the outer vessel.
In particular, the transition to \ce{H2} in Run 3 occurred at a point when the release of tritium from the salt had already plateaued, suggesting that the salt was depleted of tritium. Therefore, the additional release observed after switching to \ce{H2} likely originated from tritium retained in previous experiments, in particular Run 3 which was conducted with a much higher neutron fluence (see \cref{tab:tbr}).

To further investigate this effect, in Run 4 hydrogen was introduced from the beginning.
A significantly faster release was observed, reducing the total experimental duration to just four days (see \cref{fig:release-run-4}).
This result strongly supports the role of isotopic exchange in enhancing tritium release.
Based on these findings, dedicated campaigns are planned to systematically vary the concentration of \ce{H2} and derive empirical laws governing this behaviour.

\subsection{Tritium speciation}

\begin{figure*}[h]
  \centering
  \begin{subfigure}{0.5\textwidth}
    \includegraphics[width=\textwidth]{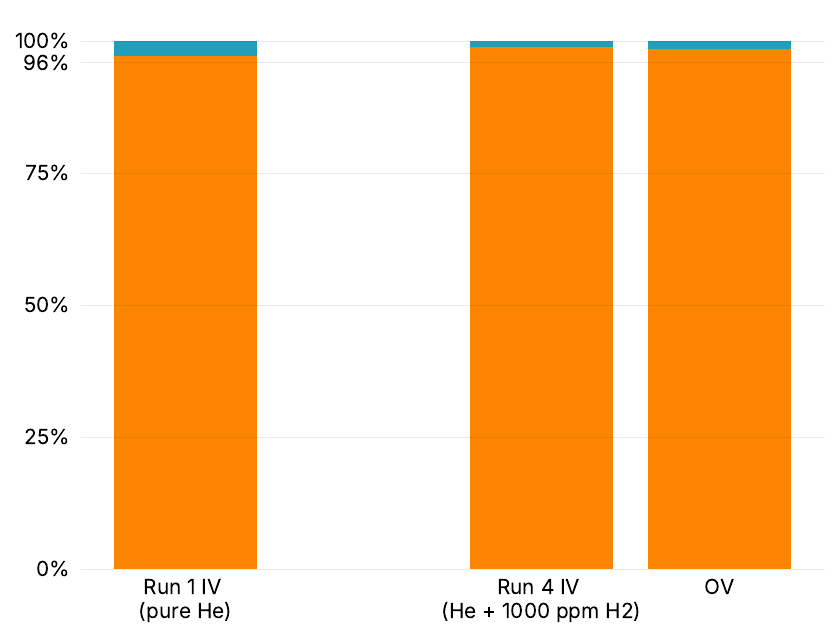}
    \caption{Comparison of the speciation fraction between Runs 1 and 4.}
    \label{fig:total-speciation}
  \end{subfigure}\\
  \begin{subfigure}{0.45\textwidth}
    \includegraphics[width=\textwidth]{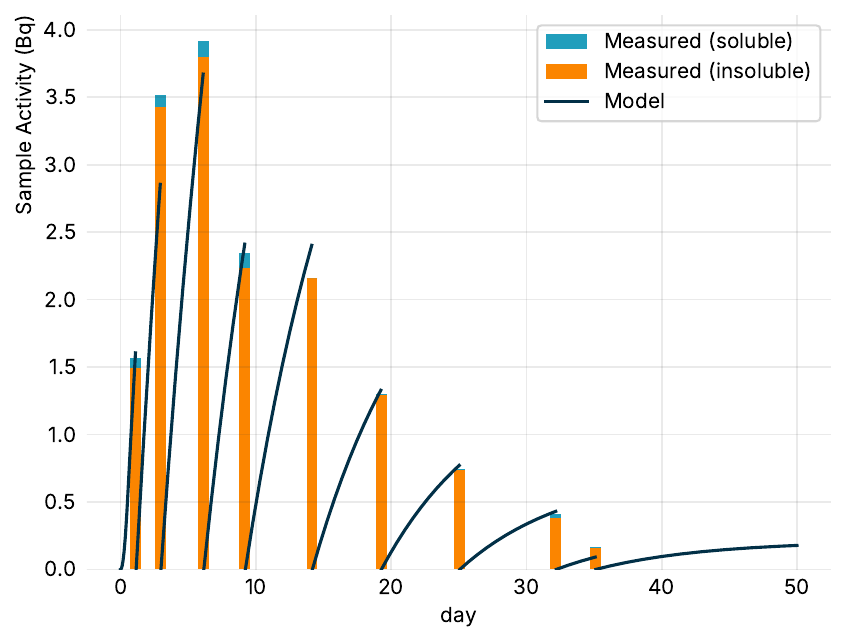}
    \caption{Temporal evolution of sample activity during Run 1 (inner vessel).}
    \label{fig:sample-speciation-run-1}
  \end{subfigure}
  \begin{subfigure}{0.45\textwidth}
    \includegraphics[width=\textwidth]{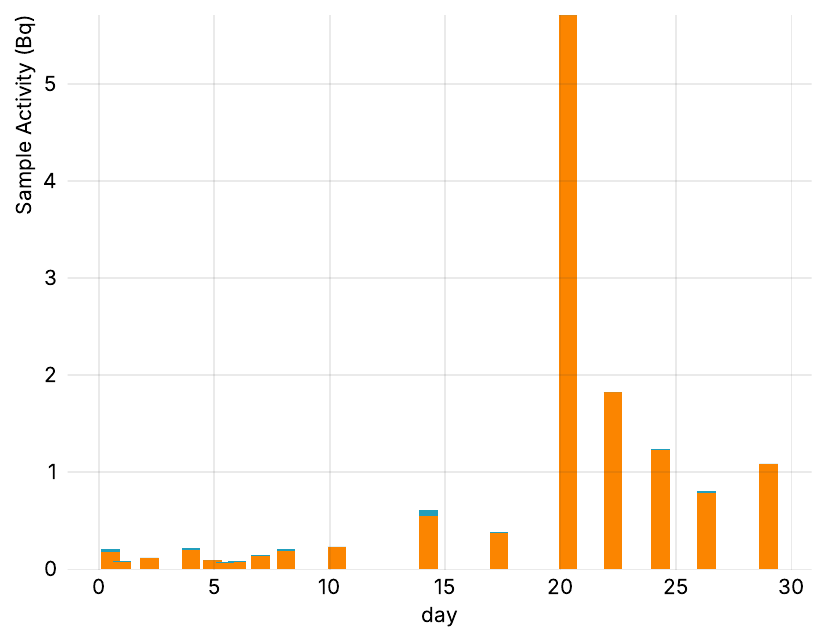}
    \caption{Temporal evolution of sample activity during Run 4 (outer vessel).}
    \label{fig:sample-speciation-run-4}
  \end{subfigure}
  \caption{Tritium speciation in the BABY 1L experiment. Orange: insoluble forms of tritium, blue: soluble forms of tritium.}
\end{figure*}

Similarly to the \SI{100}{mL} experiments, more than 96\% of the tritium was collected in water-insoluble forms (HT, \ce{T2}) regardless of the composition of the purge gas (see \Cref{fig:total-speciation}).
Therefore, the change in release dynamics (see \Cref{dynamics}) observed when adding \ce{H2} to the gas cannot be explained solely as the result of a change in the speciation of tritium or the redox state of the salt.

\section{Discussion}\label{discussion}

\subsection{Experimental uncertainties}
Several sources of experimental uncertainty affect the interpretation of the BABY 1L results. The dominant contributor is the activation foil analysis used to estimate neutron fluence.
This method relies on precise knowledge of cross sections, decay corrections, geometric positions, and detection efficiency, all of which introduce variability in the inferred neutron flux.
Additional uncertainty arises from the anisotropic nature of the sealed-tube neutron source, which is difficult to characterise fully and may cause spatial variations in flux not captured by the current foil placement.
Furthermore, while tritium release from the inner vessel can be modelled reliably, there is no clear baseline for release from the outer vessel, making it challenging to derive accurate transport parameters for that pathway.
In contrast, the uncertainty associated with tritium quantification via liquid scintillation counting is comparatively low, because of the stability of the counting system.
Overall, while LSC provides high-confidence activity measurements, the combined uncertainties in neutron flux estimation and outer vessel modelling remain the main limitations in experimental precision. \par


\subsection{Modelling uncertainties}

The neutronics model used to simulate the tritium breeding ratio (TBR) in the BABY experiment carries several sources of uncertainty.
Among these, uncertainties stemming from nuclear data—primarily cross sections—are quantifiable through established methods such as perturbation or Monte Carlo sampling (e.g., via NJOY or SANDY) \cite{fiorito_nuclear_2017, dunn_neutronics_2025}, and are expected to contribute approximately 1–5\% uncertainty to the calculated TBR.
However, these are not anticipated to be the dominant source of error.
The primary source of uncertainty is believed to arise from geometric approximations, particularly those introduced during the manual CAD-to-CSG conversion process.
These include simplified representations of complex shapes, omission of small-scale features, and the lack of inclusion of surrounding structures or experiments in the laboratory vault that may have contributed to neutron reflection during the 1L BABY experimental runs.
Additional sources of uncertainty include material composition and variation in densities, especially for high-temperature ClLiF.

\subsection{Cross irradiations from other experiments}
Beginning with Run 4, additional uncertainty in the measured tritium breeding arose from cross-irradiation by other experiments conducted in the MIT Vault Laboratory.
These concurrent activities involved external neutron sources, such as cyclotrons, operating in proximity to the BABY setup.
Although these sources are capable of producing neutrons energetic enough to breed tritium, it was not possible to accurately account for their contribution due to limited information on key parameters.
Specifically, the relative distance and orientation between the external sources and BABY, the timing and duration of their operation, the specifications of their accelerators, and the presence of neutron moderators or shielding components were either unknown or poorly documented.
Moreover, frequent changes in cyclotron target materials led to significant variability in neutron emission rates and neutron spectra, further complicating modelling efforts.
As a result, the potential influence of these cross-irradiations on tritium production could not be quantified and remains a source of uncertainty in the interpretation results (after Run 4).

\subsection{Limitations of the current model}

The current tritium release model is not sufficient to capture all the effects we have observed.
In particular, it failed to capture the isotopic exchange effects observed when switching \textit{during a run} from pure He to He + \ce{H_2}.
Other limitations stem from the 0D nature of the model assuming that quantities such as the tritium production rate $S$, tritium concentration $c_\mathrm{salt}$, are homogeneous in the salt volume.
While this may be valid for the temperature, this is unlikely to be the case for the tritium source term and tritium concentration.

Future work will therefore focus on improving the current model by 1) integrating tritium transport models with FESTIM \cite{delaporte-mathurin_festim_2024} coupled to OpenMC \cite{ROMANO201590}, 2) if needed, integrate fluid advection, and 3) capture isotopic exchange phenomena and study chemical and redox reactions happening in the molten salt, in particular at the salt-metal interface.

\section{Conclusions}

The BABY 1L campaign demonstrated significant progress in quantifying and understanding tritium breeding and release in molten salt breeder systems.
The measured Tritium Breeding Ratio (TBR) increased by a factor of six compared to the earlier \SI{100}{mL} experiment, primarily due to the larger salt volume and improved solid angle coverage.
These experimental values showed very good agreement with neutronics predictions from OpenMC, providing strong validation of the modelling approach.

Beyond breeding performance, the campaign provided new insights into tritium release dynamics.
Evidence of isotopic effects was observed, with the addition of hydrogen to the purge gas leading to a great reduction in experimental timescales from 30-60 days to just a few days.
Moreover, permeated tritium in the outer vessel stream was only detected when hydrogen was introduced, supporting the hypothesis of isotopic exchange processes enhancing release pathways.

Looking ahead, additional experiments are planned to provide more validation data by, for instance, varying the position of the neutron source.
On the modelling side, efforts will focus on improving the representation of the lag between inner and outer vessel releases observed in Runs 3 and 4, as well as advancing tritium transport simulations at the crucible level.
Incorporating diffusion, advection, and multi-material effects into these models will enable a more predictive understanding of the distinct release pathways.
Together, these developments will continue to refine the BABY platform as a benchmark for tritium breeding studies in liquid breeder systems.



\section{Data availability}

All data and scripts are available at \url{https://github.com/libra-project/baby-1l-paper} \cite{baby-paper-repo}.

\bibliographystyle{elsarticle-num} 
\bibliography{references}

\end{document}